\documentclass[a4paper,11pt]{article}
\usepackage{pos}
\usepackage{colortbl}
\usepackage{multirow}

\newcommand{\loopnumber}{l}
\newcommand{\nedges}{n_{\mathrm{edges}}}

\newcommand{\NV}{n}

\newcommand{\ND}{N_D}

\newcommand{\Divisor}{P}
\newcommand{\divisor}{p}

\newcommand{\differentialform}{\Psi}

\newcommand{\Fcomb}{F_{\mathrm{comb}}}

\newcommand{\Agen}{\Omega}

\newcommand{\absmu}{|\mu|}

\title{From geometry to phenomenology}

\author{Stefan Weinzierl}

\affiliation{PRISMA Cluster of Excellence, Institut f{\"u}r Physik, \\
Johannes Gutenberg-Universit{\"a}t Mainz, \\
D - 55099 Mainz, Germany}

\emailAdd{weinzierl@uni-mainz.de}

\abstract{
Precision calculations in quantum field theory rely very often on perturbation
theory and thus on the computation of Feynman integrals.
Feynman integrals are also fascinating objects from a mathematical point of view
and show deep connections to algebraic geometry.
Cutting-edge Feynman integrals usually have geometries of "mixed" type, for example parts of it
may correspond to a K3-surface, other parts may correspond to curves of a certain genus
and the simplest parts correspond to points.
In this talk I will discuss how to extract the geometric information from a Feynman integral
and how this information can be used to compute more efficiently Feynman integrals.
Non-trivial mixed geometries already occur in $2 \rightarrow 2$-processes at two-loops,
like Drell-Yan, Bhabha and M{\o}ller scattering.
}

\FullConference{Loops and Legs in Quantum Field Theory (LL2026)\\
12-17, April, 2026\\
Bayreuth, Germany\\}

\begin{document}
\maketitle


\section{Introduction}
High‑energy‑physics experiments -- from the LHC to future colliders -- demand
extremely precise theoretical predictions. These predictions are obtained
through perturbative quantum field theory, whose central objects are
Feynman integrals.  
Feynman integrals play a key role not only in high-energy physics, 
but also in gravitational wave physics and precision experiments at low energies.
Consequently, a large research effort is devoted to advancing the methods used to evaluate them
\cite{Chen:2020uyk,Chen:2022lzr,DHoker:2023khh,Marzucca:2023gto,
delaCruz:2024xit,Baune:2024biq,Baune:2024ber,Jockers:2024uan,
Gehrmann:2024tds,Pogel:2024sdi,Duhr:2024xsy,Gasparotto:2024bku,
DHoker:2025szl,DHoker:2025dhv,Duhr:2025ppd,Duhr:2025tdf,
Chaubey:2025adn,Duhr:2025xyy,Bargiela:2025vwl,Carrolo:2026qpu,Duhr:2026elp,Crisanti:2026rbc,Feng:2025leo,Feng:2026imq,Forner:2026vby}.
Perturbative computations are based on a few core procedures: 
the reduction of Feynman integrals and the evaluation of the corresponding master integrals. 
Both steps may demand large amounts of memory and CPU time, often becoming the bottleneck of the entire workflow. 
Consequently, any improvement in the performance of these procedures directly enlarges the set 
of feasible higher‑order calculations.  
The reduction of Feynman integrals is built on integration‑by‑parts identities \cite{Tkachov:1981wb,Chetyrkin:1981qh} 
and the Laporta algorithm \cite{Laporta:2000dsw}. 
This allows any Feynman integral to be expressed as a finite linear combination of master integrals.
Several open‑source codes implement these techniques \cite{vonManteuffel:2012np,Smirnov:2014hma,Maierhoefer:2017hyi,Wu:2023upw,Guan:2024byi}.  
The second essential task -- computing master integrals -- can be addressed 
by the method of differential equations \cite{Kotikov:1990kg,Kotikov:1991pm,Remiddi:1997ny,Gehrmann:1999as,Henn:2013pwa}, 
which is applicable both analytically and numerically. 
One first employs integration-by-parts relations to generate a system of (non-$\varepsilon$-factorised) differential equations.
This step is entirely based on linear algebra and is limited only by the available computational resources.
For analytic results one typically proceeds in two further steps:
The system is cast into an $\varepsilon$-factorised form \cite{Henn:2013pwa}.
The $\varepsilon$-factorised equations are then solved order-by-order in $\varepsilon$, 
yielding solutions in terms of iterated integrals \cite{Chen}.
The last step requires suitable boundary values.
As the boundary conditions involve fewer kinematic variables, 
they are simpler to obtain and can be recursively reduced to single‑mass vacuum integrals \cite{Liu:2022chg,Liu:2017jxz,Liu:2022mfb}.
These techniques have enabled a number of state‑of‑the‑art
calculations, including many with non‑trivial geometric structures
\cite{Henn:2025xrc,Adams:2018yfj,Honemann:2018mrb,Bogner:2019lfa,
Muller:2022gec,Giroux:2022wav,Dlapa:2022wdu,Gorges:2023zgv,
Delto:2023kqv,Jiang:2023jmk,Ahmed:2024tsg,Giroux:2024yxu,
Duhr:2024bzt,Schwanemann:2024kbg,Marzucca:2025eak,Becchetti:2025oyb,
Becchetti:2025rrz,Ahmed:2025osb,Chen:2025hzq,Coro:2025vgn,
Pogel:2022vat,Pogel:2022yat,Pogel:2022ken,Duhr:2022dxb,
Forner:2024ojj,Frellesvig:2024rea,Duhr:2025lbz,Maggio:2025jel,
Duhr:2025kkq,Pogel:2025bca,Duhr:2024uid}.
Despite these successes, two bottlenecks remain:
First of all, the integration-by-parts reduction demands significant computing resources.
This is a major practical constraint.
Secondly, there is a conceptual obstacle: 
Is it always possible to transform a given system into an $\varepsilon$-factorised form?
In this talk I report on methods, based on refs.~\cite{e-collaboration:2025frv,Bree:2025tug}, aimed at alleviating both
bottlenecks, i.e. by improving the efficiency of integration-by-parts reduction and by
presenting a systematic algorithm to obtain an $\varepsilon$-factorised differential equation.


\section{Motivating examples}

Let us start with a motivating example.
\begin{figure}
\begin{center}
\includegraphics[scale=1.0]{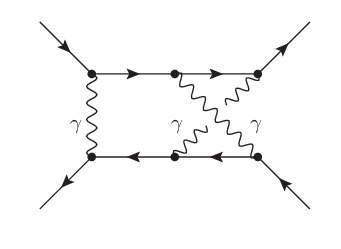}
\hspace*{10mm}
\includegraphics[scale=1.0]{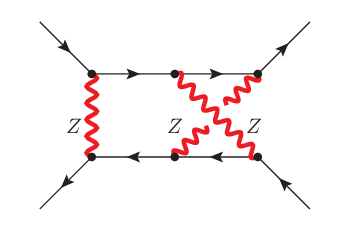}
\end{center}
\caption{
Two non-planar double box integrals.
}
\label{fig:moeller}
\end{figure}
Fig.~\ref{fig:moeller} shows two two-loop non-planar double box integrals.
In the Feynman graph on the left, three photons are exchanged, while in the Feynman graph on the right, the photons are replaced by $Z$-bosons.
Both integrals contribute within the Standard Model to two-loop electroweak corrections for 
Drell-Yan, Bhabha and M{\o}ller scattering.
The Feynman integral with the photons is well-known \cite{Tausk:1999vh}.
It depends on the Mandelstam variables $s$ and $t$.
The family of this Feynman integral has $12$ master integrals, with $2$ master integrals in the top sector and one master integral in each of the $10$ non-trivial sub-sectors.
However, the Feynman integral with the $Z$-bosons is much more involved.
It depends on the Mandelstam variables $s$, $t$ and an internal mass $m_Z$.
The family of this Feynman integral has $67$ master integrals, with $5$ master integrals in the top sector and sub-sectors with up to $8$ master integrals.

By running standard integration-by-parts reduction programs on the second example we observe that the reduction tables are significantly larger.
Part of it is inevitable, as it is a more complex example. However, we also observe that large parts are spurious and can be evaded by improving the algorithm.
To illustrate the problem we consider
the following linear system of equations:
\begin{eqnarray}
\begin{array}{rrrrcl}
 14732 \; I_1 & - 2514 \; I_2 & - 5 \; I_3 & - 7 \; I_4 & = & 0, \\
 9872 \; I_1 & - 17294 \; I_2 & + 3 \; I_3 & - 11 \; I_4 & = & 0, \\
 5068 \; I_1 & - 49336 \; I_2 & + 18 \; I_3 & - 22 \; I_4 & = & 0. \\
\end{array}
\nonumber 
\end{eqnarray}
The rank of this system is two.
If we express $I_1$ and $I_2$ in terms of the free parameters $I_3$ and $I_4$ we obtain  
\begin{eqnarray}
\label{eq_1}
 I_1 \;= \; \frac{1237}{3025750} I_3 + \frac{1229}{3025750} I_4,
 & &
 I_2 \; = \; \frac{1231}{3025750} I_3 - \frac{1223}{3025750} I_4,
\end{eqnarray}
and one immediately sees a very large denominator.
In contrast, solving the system for $I_3$ and $I_4$ while keeping $I_1$ and $I_2$ as independent parameters yields  
\begin{eqnarray}
\label{eq_2}
 I_3 \; = \; 1223 I_1 + 1229 I_2,
 & & 
 I_4 \; = \; 1231 I_1 - 1237 I_2,
\end{eqnarray}
a much simpler result that contains no large denominators. 
Our aim is to keep such denominators small.
In the context of Feynman integrals, the variables $I_1, \dots, I_4$ correspond to Feynman integrals
and the free variables to master integrals.
Large denominators correspond to lengthy polynomials in the kinematic invariants and in the dimensional‑regularisation parameter $\varepsilon$.
Such spurious polynomials often occur within sectors with a large number of master integrals.
There are some heuristic methods which try to avoid the occurrence of spurious polynomials \cite{Smirnov:2020quc,Usovitsch:2020jrk}.

Within the Laporta algorithm we 
select the free variables and the dependent variables by an order relation.
In the example above we could have used the order relation
\begin{eqnarray}
 I_1 \; > \; I_2 \; > \; I_3 \; > \; I_4 & & \mbox{(first case)},
 \nonumber \\
 I_4 \; > \; I_3 \; > \; I_2 \; > \; I_1 & & \mbox{(second case)}.
\end{eqnarray}
We see that the order relation matters for efficiency.
If we only have a linear system of equations without any further information on the unknowns, there is not much we can do about it.
However, this is not the situation we are in within the context of Feynman integral reductions.
We know that our unknowns are Feynman integrals, that they have various integral representations 
(loop momentum representation, Feynman parameter representation, Baikov representation, etc.)
and that Feynman integrals have a relation to geometry.

Within the context of this talk, the geometry associated to the left graph in fig.~\ref{fig:moeller}
consists of a set points, whereas the geometry associated to the right graph in fig.~\ref{fig:moeller} is much richer,
involving a curve of genus two in the top sector, a K3-surface in a sub-sector, five inequivalent elliptic curves in further sub-sectors and a set of points.
In general, the geometry of the maximal cut of a sector can be mixed: the five master integrals in the top sector of the example above
decompose into four master integrals
associated with the genus two curve (a variety of dimension one) and one master integral associated with a point (a variety of dimension zero).


\section{Getting technical}

In a nutshell, we perform two steps:
In the first step we run the Laporta algorithm with an order relation motivated by geometry to obtain a basis of master integrals $J$.
We observe that an order relation motivated by geometry reduces or avoids spurious polynomials in the denominator.
Furthermore, the differential equation for $J$ is compatible with a filtration, the precise meaning will be given below.
Thirdly, this step is entirely rational.

In the second step, we rotate the basis $J=R_2 K$ through a rotation matrix $R_2$ to a new basis $K$, such that the differential equation for the basis $K$ is
$\varepsilon$-factorised:
\begin{eqnarray}
 d K & = & \varepsilon A\left(x\right) K.
\end{eqnarray}
For a filtration-compatible differential equation this can always be done.
The rotation matrix $R_2$ can be transcendental.

There are two branches of mathematics, which are relevant: Twisted cohomology and Hodge theory.
Feynman integrals may have ultraviolet and infrared divergences
and we use dimensional regularisation to regulate them.
In a mathematical language this translates to twisted cohomology.
Hodge theory provides an abstraction, which treats all geometries of finite Feynman integrals in one single framework.
Parts of Hodge theory carry over from ordinary cohomology (e.g. finite Feynman integrals) to twisted cohomology (e.g. dimensionally regulated Feynman integrals).
At the end of the day, it boils down for the order relation to count residues and poles.

\begin{figure}
\begin{center}
\includegraphics[scale=1.0]{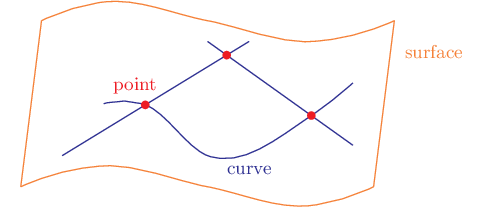}
\end{center}
\caption{
A mixed geometry:
Inside a surface of dimension two there can be
curves of dimension one and points of dimension zero.
}
\label{fig:mixed_geometry}
\end{figure}

We may write down integration-by-parts identities with respect to
any integral representation. Examples are the loop momentum representation
or the Baikov representation.
The integration-by-parts identities may look different in the various integral representations, but contain the same information.
As we are mainly interested in the order within one sector, we work in the following with the 
maximal cut of a loop-by-loop Baikov representation.
We define the Baikov polynomials $p_i(z)$ by
\begin{eqnarray}
 \int\limits_{{\mathcal C}_{\mathrm{maxcut}}} \prod\limits_{r=1}^{\loopnumber} \frac{d^Dk_r}{i \pi^{\frac{D}{2}}} 
 \frac{1}{\prod\limits_{j=1}^{\nedges} \sigma_j}
 & \sim & 
 \int d^{\NV}z \;
 \prod\limits_{i \in I_{\mathrm{all}}} \left[ \divisor_i\left(z\right) \right]^{\alpha_i},
\end{eqnarray}
where the $\sigma_j$'s denote the inverse propagators.
The exponents $\alpha_i$ are always of the form
\begin{eqnarray}
 \alpha_i \; = \; 
 \frac{1}{2} \left( a_i + b_i \varepsilon \right),
 & \mbox{with} &
 a_i,b_i \; \in \; {\mathbb Z}.
\end{eqnarray}
Instead of working in the affine chart $z=(z_1,\dots,z_{\NV})$, it is convenient to go to projective space ${\mathbb C}{\mathbb P}^{\NV}$
with homogeneous coordinates $[z_0:z_1:\dots:z_{\NV}]$.
We denote by $P_i$ the homogenisation of $p_i$.
To preserve homogeneity, we introduce $P_0=z_0$ with an appropriate exponent $\alpha_0 = \frac{1}{2}(a_0+b_0\varepsilon)$.
We define $I^0_{\mathrm{odd}}$ as the set of indices for which $a_i$ is odd
and $I^0_{\mathrm{even}}$ as the set of indices for which $a_i$ is even.
Within twisted cohomology, we may always move integer powers of the Baikov polynomials
between the twist function and the rational differential $\NV$-form.
We can therefore define a ``minimal'' twist function,
by requiring $a_i \in \{-1,0\}$ for all $i$:
\begin{eqnarray}
 U\left(z_0,z_1,\dots,z_{\NV}\right)
 & = &
 \prod\limits_{i \in I_{\mathrm{odd}}^0} \Divisor_i^{-\frac{1}{2}+\frac{1}{2} b_i \varepsilon}
 \prod\limits_{j \in I_{\mathrm{even}}^0} \Divisor_j^{\frac{1}{2} b_j \varepsilon}.
\end{eqnarray}
We then study the integrand
\begin{eqnarray}
 \differentialform_{\mu_0 \dots \mu_{\ND}}\left[Q\right]
 & = &
 \mbox{prefactor}
 \cdot 
 U
 \; \cdot
 \frac{Q}{\prod\limits_{i \in I_{\mathrm{all}}^0} \Divisor_i^{\mu_i}}
 \cdot \;
 \mbox{standard $\NV$-form}.
\end{eqnarray}
The even and the odd polynomials play different roles.
If the denominator contains an even polynomial, we may take a residue, 
thereby reducing the problem to a simpler one of dimension $(\NV-1)$.
The odd polynomials define a geometry
\begin{eqnarray}
\label{def_hypersurface_square_free}
 y^2 - \prod\limits_{i \in I^0_{\mathrm{odd}}} \Divisor_i\left(z\right) & = & 0
\end{eqnarray}
in an appropriately chosen weighted projective space.
On the maximal cut we have in general a mixed geometry: Inside the geometry of dimension $\NV$ defined by eq.~(\ref{def_hypersurface_square_free})
there can be distinguished sub-geometries of dimension $(\NV-1)$, which in turn might have distinguished sub-sub-geometries of dimension $(\NV-2)$, etc.
This is sketched in fig.~\ref{fig:mixed_geometry}.
For an integrand $\differentialform_{\mu_0 \dots \mu_{\ND}}[Q]$ we now define the following numbers:
\begin{eqnarray}
 o & = & \mbox{pole order},
 \nonumber \\
 r & = & \mbox{number of consecutive non-zero residues},
 \nonumber \\
 w & = & \NV + r
 \nonumber \\
 \absmu & = & \mu_0+\dots+\mu_{\ND}.
\end{eqnarray}
The quantity $w$ denotes the Hodge weight. We further set $a=-w$, if $\differentialform_{\mu_0 \dots \mu_{\ND}}[Q]$ is the pre-image of a master integrand of a sub-problem
localised on $\Divisor_i=0$ with $i \in I_{\mathrm{even}}^0$ and $a=0$ otherwise.
With these preparations we may now give the geometric order relation: We order the Feynman integrals by
\begin{eqnarray}
 \left( \nedges, \mathrm{sector} \; \mathrm{id}, a,w,o,|\mu|,\dots \right)
\end{eqnarray}
where
$(\nedges, \mathrm{sector} \; \mathrm{id})$ orders first by sectors,
followed by $(a,w,o,|\mu|)$, which gives the dominant order within a sector and
the dots stand for further criteria needed to distinguish inequivalent integrals.

\begin{figure}
\begin{center}
\includegraphics[scale=0.5]{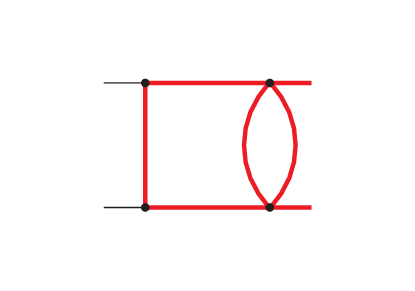}
\hspace*{5mm}
\includegraphics[scale=0.5]{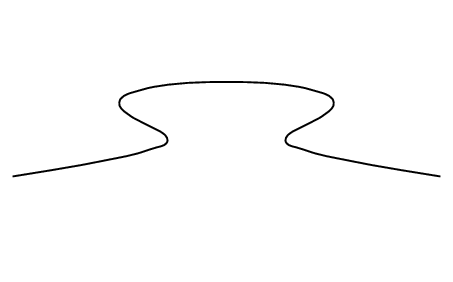}
\hspace*{5mm}
\includegraphics[scale=0.5]{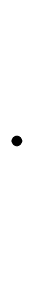}
\hspace*{5mm}
\includegraphics[scale=0.5]{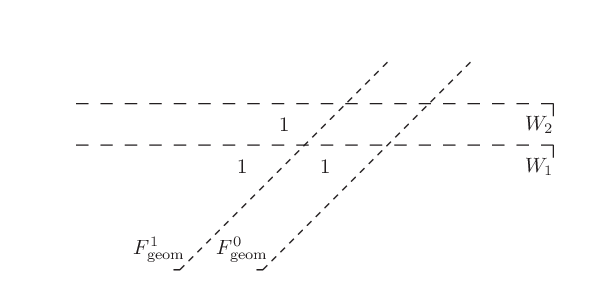}
\end{center}
\caption{
Left part: The Feynman graph. Red lines correspond to a particle of mass $m$, black lines to
a massless particle.
Middle part: The geometries consist of an elliptic curve and a point.
Right part: The Hodge-like diagram.
}
\label{fig:sector79}
\end{figure}

As an example we consider an elliptic sector, contributing to $pp\rightarrow t \bar{t}$, with three master integrals.
The Feynman graph is shown in the left part of fig.~\ref{fig:sector79}.
As geometry we have an elliptic curve and a point, sketched in the middle part of fig.~\ref{fig:sector79}.
We may decompose the three-dimensional vector-space of master integrals into three sub-spaces as shown in the right part of fig.~\ref{fig:sector79}.

We observe that the basis $J$ obtained from the Laporta algorithm with the geometric order relation has the property that on the maximal cut
we have
\begin{eqnarray}
\label{F_compatible}
 d J
 \; = \;
 A\left(\varepsilon,x\right) J,
 & &
 A_{ij}\left(\varepsilon,x\right)
 \; = \;
 \sum\limits_{k=-(\absmu_i-\absmu_j)}^1
 \varepsilon^k A^{(k)}_{ij}\left(x\right).
\end{eqnarray}
We call such a differential equation an $\Fcomb^\bullet$-compatible differential equation,
where the filtration $\Fcomb^\bullet$ is defined by 
\begin{alignat}{2}
 \differentialform_{\mu_0 \dots \mu_{\ND}}[Q] & \in \Fcomb^{p} \Agen^{\NV} & \quad \mbox{if} & \quad \NV-\absmu \ge p.
\end{alignat}
From such a form we may always construct algorithmically a matrix $R_2$,
which leads to a basis $K=R_2^{-1}J$, such that 
the differential equation for $K$ is in $\varepsilon$-factorised form.
We illustrate this with an example featuring three non-trivial components in the filtration:
\begin{eqnarray}
 \emptyset = \Fcomb^3 \Agen^{\NV} \subseteq \Fcomb^2 \Agen^{\NV} \subseteq \Fcomb^1 \Agen^{\NV} \subseteq F^0 \Agen^{\NV} = \Agen^{\NV}.
\end{eqnarray}
In this case we can write for the matrix $A$ appearing in $dJ=AJ$
\begin{eqnarray}
 A & = & B^{(1)} + B^{(0)} + B^{(-1)} + B^{(-2)},
\end{eqnarray}
where
\begin{align}
\label{example_B}
 B^{(1)}
 &= 
 \left( \begin{array}{rrr}
 \varepsilon A^{(1)}_{11} & \varepsilon A^{(1)}_{12} & 0 \\ 
 \varepsilon A^{(1)}_{21} & \varepsilon A^{(1)}_{22} & \varepsilon A^{(1)}_{23}  \\ 
 \varepsilon A^{(1)}_{31} & \varepsilon A^{(1)}_{32} & \varepsilon A^{(1)}_{33}  \\ 
 \end{array} \right),
 &
 B^{(0)}
 & = 
 \left( \begin{array}{ccc}
 0 & 0 & 0 \\ 
 0 & 0 & 0  \\ 
 \cellcolor{green} A^{(0)}_{31} & 0 & 0   \\ 
 \end{array} \right),
 \nonumber \\
 B^{(-1)}
 & =
 \left( \begin{array}{ccc}
 0 & 0 & 0 \\ 
 \cellcolor{magenta} A^{(0)}_{21} & 0 & 0  \\ 
 \cellcolor{magenta} \frac{1}{\varepsilon} A^{(-1)}_{31} & \cellcolor{cyan} A^{(0)}_{32} & 0   \\ 
 \end{array} \right),
 &
 B^{(-2)}
 & = 
 \left( \begin{array}{ccc}
 \cellcolor{yellow} A^{(0)}_{11} & 0 & 0 \\ 
 \cellcolor{yellow} \frac{1}{\varepsilon} A^{(-1)}_{21} & \cellcolor{orange} A^{(0)}_{22} & 0  \\ 
 \cellcolor{yellow} \frac{1}{\varepsilon^2} A^{(-2)}_{31} & \cellcolor{orange} \frac{1}{\varepsilon} A^{(-1)}_{32} & \cellcolor{red} A^{(0)}_{33}   \\ 
 \end{array} \right).
\end{align}
We remove unwanted terms step-by-step, starting with $B^{(-2)}$, then $B^{(-1)}$, and finally $B^{(0)}$.
Within each $B^{(k)}$ we proceed column-wise, starting with column one.
In general, the matrix $R_2$ will involve transcendental functions. 
These additional functions satisfy differential equations, which are similar, but simpler than the ones for the Feynman integrals of interest.

The algorithm extends beyond the maximal cut. 
For sub-sectors we do not need to require that the $\varepsilon$-dependence of the terms 
in the non-diagonal blocks is given by a Laurent polynomial in $\varepsilon$.
The algorithm can handle a rational dependence in $\varepsilon$: 
One first performs a partial fraction decomposition in $\varepsilon$ and then
removes any term which is not proportional to $\varepsilon^1$.


\section{Efficiency improvement}

\begin{figure}
\begin{center}
\includegraphics[scale=0.6]{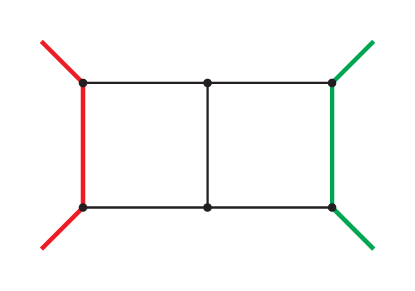}
\end{center}
\caption{
The H-graph.
Red lines correspond to a particle of mass $m_1$, green lines to a particle of mass $m_2$
and black lines to
a massless particle.}
\label{fig:H-graph}
\end{figure}

As an example for the efficiency improvement obtained with this algorithm we consider the 
H-graph in quantum field theory \cite{Kreer:2024zzf}, shown in fig.~\ref{fig:H-graph}.
This graph depends on $s$, $t$, $m_1^2$ and $m_2^2$.
A measure, which is independent of the implementation of the algorithm, is the required disk space for the differential equation.
We find for the required disk space
\begin{center}
\begin{tabular}{|l|r|}
\hline
 standard order relation & 18433 kB \\
 geometric order relation & 112 kB \\
 $\varepsilon$-factorised & 9 kB \\
\hline
\end{tabular}
\end{center}
In total, we roughly observe a reduction in disk space by a factor of $2000$.
The basis $J$, obtained after step $1$, reduces the required disk space roughly by a factor of $200$.
This is explained by the avoidance of spurious polynomials in the denominator of our algorithm.
A further factor of $10$ is obtained by going from the basis $J$ to the $\varepsilon$-factorised basis $K$, this is explained by the avoidance of 
copies of similar or related expressions at different orders in $\varepsilon$.


\section{Conclusions}

In this talk I argued that 
a well-chosen order relation in integration-by-parts reductions can significantly improve the efficiency of the reduction by
reducing or avoiding spurious polynomials in the denominator.
We observed that an order relation motivated by the underlying geometry of Feynman integrals performs significantly better
than order relations used up to now.

I also presented a systematic algorithm to obtain a $\varepsilon$-factorised differential equation.
The algorithm consists of two steps:
In the first step one obtains
an intermediate basis $J$ directly from the order relation.
In a second step on rotates to a basis $K$ to obtain the $\varepsilon$-factorised differential equation.
This two-step procedure separates the rational and the transcendental part: Step $1$ is entirely rational, while
step $2$ may introduce transcendental functions.
The algorithm is based on twisted cohomology and uses
ideas from Hodge theory.


{\footnotesize
\bibliography{/home/stefanw/notes/biblio}
\bibliographystyle{h-physrev5}
}

\end{document}